\documentclass[twocolumn]{aa}
\usepackage{graphicx}
\usepackage{txfonts}
\begin{document}

\title{Four new wide binaries among exoplanet host stars.\thanks{Based on observations obtained
on La Silla in ESO programs 70.C-0116(A), 71.C-0140(A) ,72.C-0571(B), 73.C-0103(A) and on Mauna Kea
in UKIRT program U/02A/16.}}

\author{M. Mugrauer \inst{1} \and R. Neuh\"auser \inst{1} \and A. Seifahrt \inst{1}$^{,}$\inst{2} \and T. Mazeh \inst{3}
\and E. Guenther \inst{4}}

\offprints{Markus Mugrauer, markus@astro.uni-jena.de}

\institute{Astrophysikalisches Institut, Universit\"at Jena, Schillerg\"a{\ss}chen 2-3, 07745 Jena, Germany \\
\and European Southern Observatory, Karl-Schwarzschild-Str. 2, 85748 Garching, Germany \\
\and Tel Aviv University, Tel Aviv 69978, Israel \\
\and Th\"uringer Landessternwarte Tautenburg, Sternwarte 5, 07778 Tautenburg, Germany \\}

\date{Received ??? ; Accepted ???}

\abstract{In our ongoing survey for wide (sub)stellar companions of exoplanet host stars we have
found 4 new co-moving stellar companions of the stars HD\,114729, HD\,16141, HD\,196050 and
HD\,213240 with projected separations from 223 up to 3898\,AU. The companionship of HD\,114729\,B,
HD\,196050\,B and HD\,213240\,C is confirmed by photometry and spectroscopy, all being early M
dwarfs. The masses of the detected companions are derived from their infrared JHK magnitudes and
range between 0.146 and 0.363\,$M_{\sun}$. Our first and second epoch observations can rule out
additional stellar companions around the primaries from $\sim$\,200 up to $\sim$\,2400\,AU
(S/N=10). In our survey we have found so far 6 new binaries among the exoplanet host stars.
According to these new detections, the reported differences between single-star and binary-star
planets with orbital periods short than 40 days remain significant in both the mass-period and
eccentricity-period distribution. In contrast, all exoplanets with orbital periods longer than 100
days tend to display similar distributions.

\keywords{Stars:low-mass, planetary systems  --stars individual: HD\,16141 , HD\,114729 ,
HD\,196050 , HD\,213240}}

\maketitle

\section{Introduction}
Precise radial velocity (RV) search campaigns have found more than 130 exoplanets
\footnote{summarized at http://www.obspm.fr/encycl/encycl.html or http://exoplanets.org/} orbiting
G to M stars most of them in the solar neighborhood. Most exoplanets in the northern sky were
detected, with the radial velocity method by the California \& Carnegie planet search (Marcy \&
Butler 1996) and the ELODIE northern extrasolar planet search (Mayor \& Queloz 1995). In the
southern sky exoplanets are searched for by the Anglo-Australian planet search team (Tinney et al.
2001), in the CORALIE survey (Queloz et al. 2000) and by the recently started HARPS project (Pepe
et al. 2004), among others.

Some of those exoplanets orbit stars that are themselves members of a multiple stellar system
(binaries e.g. HD\,19994 or triples e.g. 16\,Cyg). The planets in those systems are interesting
objects because they provide the possibility to study the effect of stellar multiplicity on planet
formation, long-time stability and evolution of planetary orbits. Some authors have started to
compare orbital properties of known planets in binaries with those of planets found in single
stellar systems. There seem to be differences in the mass-period or eccentricity-period
distributions between planets of single stars and planets of binaries as pointed out by e.g.
Zucker\& Mazeh (2002) or Eggenberger et al. (2004). However the number of known multiple stellar
systems that harbor exoplanets is rather small (e.g. 15 shown by Eggenberger et al. (2004)). Hence,
statistical differences are sensitive to changes in the sample, either retracted planet detections,
e.g. Desidera et al. (2004), or newly found systems with exoplanets as presented here. Furthermore
stars listed as binaries in some double star catalogs cannot be confirmed by follow-up observations
(see e.g. Sec.\,3.3). Only systematic search programs for (sub)stellar companions can reveal the
real fraction of wide ($\ge$ few arcsec) visual companions among all stars known to harbor planets.
This is the first important step which must be done before analyzing statistical differences.

Several groups have already searched for (sub)stellar companions orbiting RV planet host stars with
adaptive optics and found either companions around a few of them, like HD\,114762 and $\tau$ Boo
(Patience et al. 2004), Gl\,86 (Els et al. 2001 and Mugrauer\,\&\,Neuh\"auser 2005) or could
exclude additional close faint companions (e.g. Macintosh et al. 2003). However, an interesting
regime of companions with separations up to $\sim$\,1000\,AU is not accessible to those searches,
due to their small field of view (FOV). Lowrance et al. (2002) presented a first new wide (750\,AU)
low-mass stellar companion (m$\sim$ 0.2M\,$_{\sun}$) which was detected in the digitized plates of
the Palomar Obervatory Sky Survey. However, so far the whole sample of exoplanet host stars has not
been surveyed homogeneously for such wide companions and therefore we initiated a systematic search
campaign for wide faint companions of all exoplanet host stars. We use relatively large fields of
view of up to 150\,arcsec. Companions are detected by common proper motion first, and then are
confirmed by photometry and spectroscopy (consistency of the apparent magnitude, color and spectral
type of the companion at the distance of the primary).

Our effort already yielded two new wide stellar companions to the exoplanet host stars HD\,75289\,A
(Udry et al. 2000) and HD\,89744\,A (Korzennik et al. 2000). HD\,75289\,B is a low-mass stellar
companion (0.135\,$M_{\sun}$) at a projected separation of 621\,AU, detected with SofI\footnote{Son
of Isaac} at the ESO New Technology Telescope (NTT) (see Mugrauer et al. 2004a). The companionship
of HD\,89744\,B, proposed by Wilson et al. (2001), was confirmed with astrometry by
UFTI\footnote{UKIRT Fast-Track Imager} at the United Kingdom Infrared Telescope (UKIRT). In this
case the companion mass ranges between 0.072 and 0.081\,$M_{\sun}$, at an orbit with a projected
separation of 2456\,AU (see Mugrauer et al. 2004b).

In this work we present 4 new wide low-mass stellar companions which we detected in our ongoing
survey orbiting HD\,114729, HD\,196050, HD\,213240 and HD\,16141. Section\,2 describes the
observations and reduction. In Section 3 to 5 we present astrometric, photometric and spectroscopic
evidence for companionship. Section 6 summarizes the properties of the components of the newly
found four binaries as well as their exoplanets and presents the detection limits of our
observations.

\section{Imaging, data reduction and calibration}

All observations of our program are carried out in near infrared H band. Low-mass (sub)stellar
companions are much brighter in the IR than in optical bands, hence detectable with a 4\,m class
telescope. In our survey we use two telescopes equipped with IR cameras. Imaging of targets in the
northern sky is done with the 3.8\,m UKIRT on Mauna Kea and the UFTI IR camera, equipped with a
1024x1024 HgCdTe-detector with a pixelscale of $\sim$91\,mas, i.e. 93x93\,arcsec FOV. For the
southern sky we use the 3.58\,m NTT and its SofI IR camera, located at La Silla, Chile. SofI
includes a 1024x1024 HgCdTe-detector with a pixelscale of $\sim$144\,mas, i.e. 147x147\,arcsec FOV.
All exoplanet host stars are relatively bright targets, much brighter than their possible
companions. Hence, the bright primaries saturate the detectors, which makes a companion search
impossible close to the primaries. To reduce this contamination we always use individual exposure
times as short as possible, 1.2\,s with SofI and 4\,s with UFTI. Several of those images are taken
at the same position and are stacked together by averaging. Then the telescope is moved to another
position where the same procedure is repeated. Finally all images are flatfielded and combined to
the final image with the ESO package ECLIPSE\footnote{ESO C Library for an Image Processing
Software Environment} which also provides the background subtraction. All images of an observing
run were astrometrically calibrated by comparing the positions of the detected objects in our
images with positions at the 2MASS\footnote{2 Micron All Sky Survey} point source catalogue, which
contains accurate astrometry of objects brighter than 15.2\,mag in H (S/N$>$5). In
Table\,\ref{calibration} we show the pixelscale as well as the derived detector orientation for
SofI and UFTI observing runs from which data are presented in this work.

\begin{table} [htb]
\caption{The astrometrical calibration of all observing runs for which data are shown in this
paper. The pixelscale $PS$ and the detector position angle $PA$ with their uncertainties are
listed. The detector is tilted by $PA$ from north to west.}
\begin{center}
\begin{tabular}{c|c|c|c}
\hline\hline
instrument & epoch & PS[$''$] & PA [$^{\circ}$]\\
\hline

UFTI & 11/02 & 0.09098$\pm$0.00043 &  \,\,\,0.770$\pm$0.087\\
UFTI & 10/03 & 0.09104$\pm$0.00030 &  \,\,\,0.711$\pm$0.083\\
\hline
SofI$_{small}$ & 12/02 & 0.14366$\pm$0.00016  & 90.069$\pm$0.041\\
SofI$_{small}$ & 06/03 & 0.14365$\pm$0.00013  & 90.069$\pm$0.032\\
SofI$_{small}$ & 07/04 & 0.14356$\pm$0.00011  & 90.047$\pm$0.024\\
\hline
SofI$_{large}$ & 07/04 & 0.28796$\pm$0.00032  & 90.176$\pm$0.045\\

\hline\hline
\end{tabular}
\label{calibration}
\end{center}
\end{table}

\section{Astrometry}

In principle, all objects located close to an exoplanet host star could be real companions, i.e.
they are all companion-candidates and therefore must be examined. However, most of these candidates
will emerge as ordinary background stars randomly located close to but far behind the host star. A
real companion shares the proper motion of its primary star because its orbital motion is much
smaller than the common proper motion. Hence, astrometry is an effective method to detect and
distinguish real companions from none moving background stars.

This section presents astrometric detection of four new wide co-moving companions. The
astrometric test for companionship will be confirmed by photometry and spectroscopy in the
following sections.

\subsection{HD\,114729}

\begin{figure} [ht] \resizebox{\hsize}{!}{\includegraphics{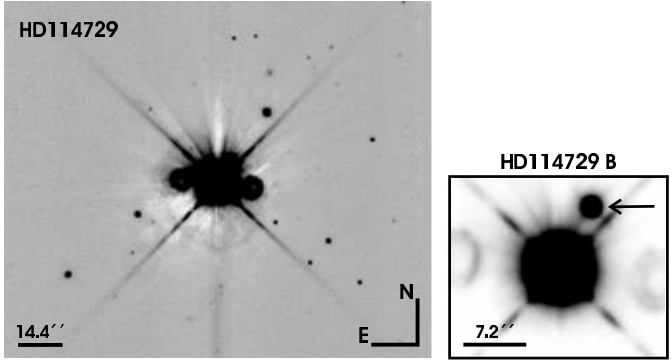}}
\caption{Left: The SofI small field image of the planet host star HD\,114729 (central bright star
with diffraction spikes and reflections), taken in June 2003 in the H band. The total integration
time is 10\,min. Several faint companion-candidates can be seen. Right: Magnified image of the
central part of the FOV with different cuts. A companion-candidate, located $\sim$\,8\,arcsec
northwest of the central star, is visible.} \label{hd114image}
\end{figure}

\begin{figure} [ht] \resizebox{\hsize}{!}{\includegraphics{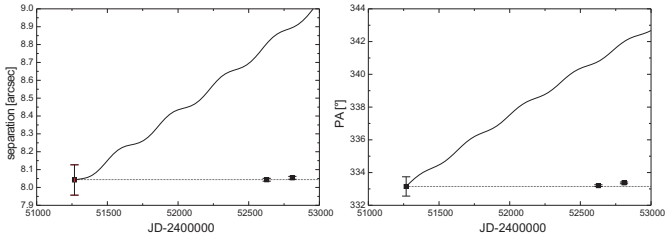}}
\caption{The separation and the position angle between HD\,114729\,B and its primary in the 2MASS
image (epoch 04/99) and our SofI images from epoch 12/02 and 06/03. If HD\,114729\,B is a
non-moving background star, both parameters must change due to the well-known proper and
parallactic motion of the primary star which is illustrated with a solid line. However, the
measured separation and position angle are constant. With the given astrometric uncertainties the
background hypothesis can be rejected at a 10\,$\sigma$ level in separation and at a 15\,$\sigma$
level in position angle. Hence, HD\,114729\,B is clearly co-moving with the exoplanet host star
HD\,114729\,A.} \label{hd114astro1}
\end{figure}

In our ongoing survey, HD\,114729 was observed in December 2002 (epoch 12/02) and in epoch 06/03
with SofI. We found a companion-candidate only 8\,arcsec northwest of the exoplanet host star, also
detected by 2MASS in epoch 04/99. We used the \textsl{Starlink} software \textsl{GAIA} and its
object detection routine to measure the separation and the position angle of the
companion-candidate relative to the primary in our SofI images. The astrometry of both objects in
the 2MASS image is provided by the 2MASS point source catalogue. If the candidate is a background
object, it would have negligible proper motion and we would expect a change in separation and
position angle due to the proper and parallactic motion of the exoplanet host star. This expected
relative motion can be derived from the Hipparcos data of the primary
($\mu_{Ra}$=-202.11$\pm$0.39\,mas/yr, $\mu_{Dec}$=-308.49$\pm$0.70\,mas/yr and
$\pi$=28.57$\pm$0.97\,mas). However, the derived separation and position angle are constant in all
three epochs. Hence, the detected companion-candidate is clearly co-moving and will be denoted as
HD\,114729\,B (see Fig.\,\ref{hd114astro1} and Tab.\,\ref{astrometry}).

\begin{figure} [ht] \resizebox{\hsize}{!}{\includegraphics{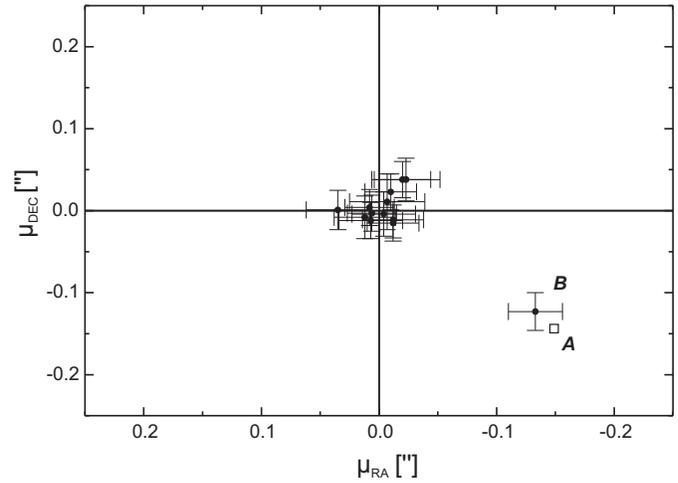}}
\caption{The derived proper motion of all companion candidates detected in our NTT images
(S/N$>$10) from epochs 12/02 and 06/03. The expected proper motion of the primary is derived from
Hipparcos data and is shown as a small square in the plot (A). HD\,114729\,B clearly shares the
proper motion of the exoplanet host star. All other detected companion-candidates prove to be
non-moving background stars.} \label{hd114astro2}
\end{figure}

Our first epoch SofI image shows several faint stars (S/N$=$10 at H\,$\sim$\,18\,mag) that are
invisible in the less sensitive 2MASS image (S/N=10 at H\,$\sim$\,14.4\,mag). To derive the proper
motion of these faint objects as well as to confirm that HD\,114729\,B is co-moving with the
exoplanet host star, we obtained a second epoch SofI image only half a year later, in epoch 06/03.
The astrometry of all companion-candidates (S/N$>$10) is measured with \textsl{ESO MIDAS} using
Gaussian-fitting. All detected objects but HD\,114729\,B have proper motions which are negligible
within the astrometric uncertainties (20 to 30\,mas), hence are non-moving background stars. The
derived proper motion of HD\,114729\,B is consistent with the expected value for a co-moving
companion for the given epoch difference (see Fig.\,\ref{hd114astro2}).

\subsection{HD\,196050}

\begin{figure} [ht] \resizebox{\hsize}{!}{\includegraphics{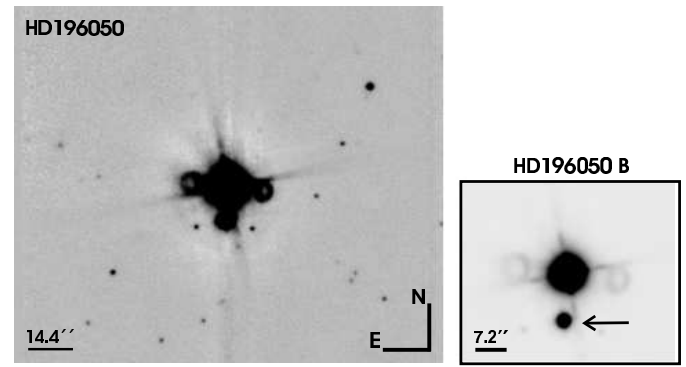}}
\caption{Left: The SofI small field image of the planet host star HD\,196050, obtained in epoch
06/03 in H band. The total integration is 10\,min. Several faint companion-candidates are detected.
Right: Image of the central region of the field. The co-moving companion, HD\,196050\,B is located
$\sim$\,11\,arcsec south of its primary star.} \label{hd196image}
\end{figure}

\begin{figure} [ht] \resizebox{\hsize}{!}{\includegraphics{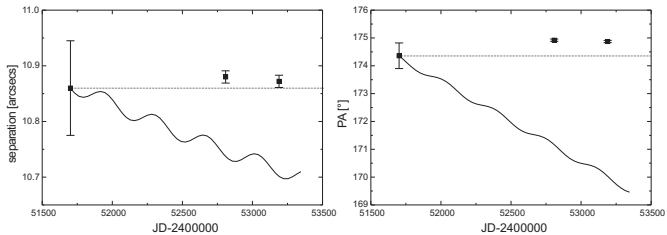}}
\caption{Separation and position angle of HD\,196050\,B to its primary for the 2MASS image (epoch
06/00) and our SofI images (epoch 06/03 and 07/04). The expected variation of both parameters if
HD\,196050\,B was a non-moving object is illustrated with a solid line. The separation and the
position angle are constant within the astrometric uncertainty, as it is expected for a co-moving
companion and significantly (11$\sigma$ in position angle) different from the expected values for a
non-moving HD\,196050\,B.} \label{hd196astro1}
\end{figure}

In June 2003 and July 2004 we observed HD\,196050 with SofI. Figure\,\ref{hd196image} shows a faint
star only $\sim$\,11\,arcsec south of the exoplanet host star, also detected by the 2MASS survey in
June 2000. As described in the previous section we compared the relative position of the faint
objects in 2MASS and our images (see Fig.\,\ref{hd196astro1} and Table\,\ref{astrometry}). The
detected companion-candidate emerges as a co-moving companion and will be denoted as HD\,196050\,B.
As expected for a co-moving companion, the separation from the primary as well as its position
angle is constant within the astrometric uncertainty for all given epochs. If HD\,196050\,B is not
co-moving, the position angle must significantly change (11$\sigma$) according to Hipparcos data of
HD\,196050\,A ($\mu_{Ra}$=-190.97$\pm$0.71\,mas/yr, $\mu_{Dec}$=-64.27$\pm$0.57\,mas/yr and
$\pi$=21.31$\pm$0.91\,mas).

\begin{figure} [ht] \resizebox{\hsize}{!}{\includegraphics{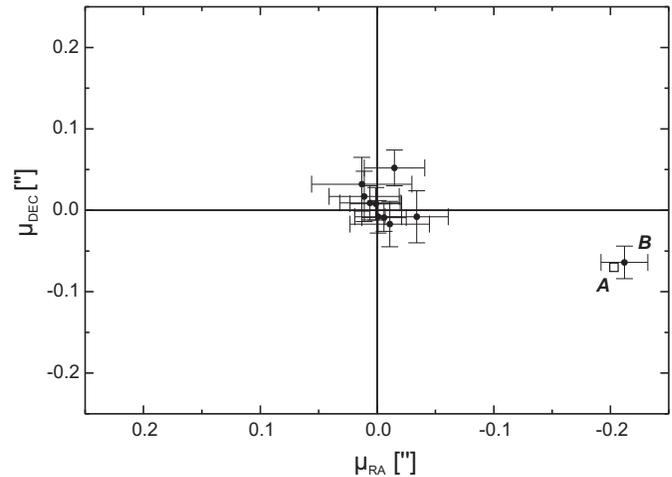}}
\caption{The derived proper motion of all detected companion candidates (S/N$>$10) shown in
Fig.\ref{hd196image} for our two SofI images taken in epoch 06/03 and 07/04. The expected proper
motion of the primary is derived with the well-known proper and parallactic motion of the primary
(from Hipparcos data) and is shown as a small square in the plot (A). HD\,196050\,B is clearly
co-moving to the exoplanet host star, but all other detected companion-candidates emerge as
non-moving background stars.} \label{hd196astro2}
\end{figure}

The first epoch SofI image shows several faint companion-candidates, which are all well detected
with S/N$\ge$10, hence accurate astrometry is available for them. In July 2004 we obtained the
second epoch SofI image and we determined the proper motion of all detected companion-candidates as
described in the previous section. The derived proper motion of HD\,196050\,B is consistent with
its primary star HD\,196050\,A (see Fig.\,\ref{hd196astro2}) as it is expected for a co-moving
companion. Furthermore, with the accurate NTT astrometry we can clearly rule out all other stars as
co-moving with the exoplanet host star.

\subsection{HD\,213240}

\begin{figure} [ht] \resizebox{\hsize}{!}{\includegraphics{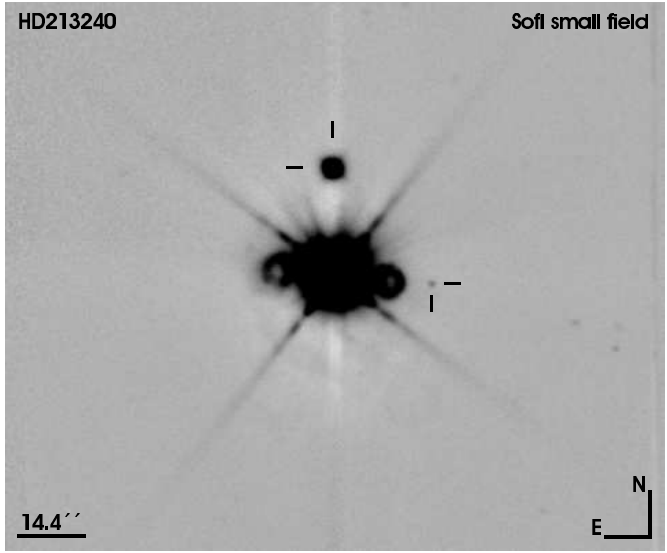}}
\caption{The SofI small field image of the exoplanet host star HD\,213240. It was taken in July
2004 in H band with a total integration of 10\,min. Only two companion-candidates both are marked
in the image are detected with S/N$\geq$10 and offer a sufficient astrometry to check common proper
motion with the bright primary. The separation of both companion-candidates to HD\,213240 varies
over time as expected from the well-known proper and parallactic motion of the exoplanet host star,
hence they are non-moving background stars.} \label{hd213imagesmall}
\end{figure}

\begin{figure} [ht] \resizebox{\hsize}{!}{\includegraphics{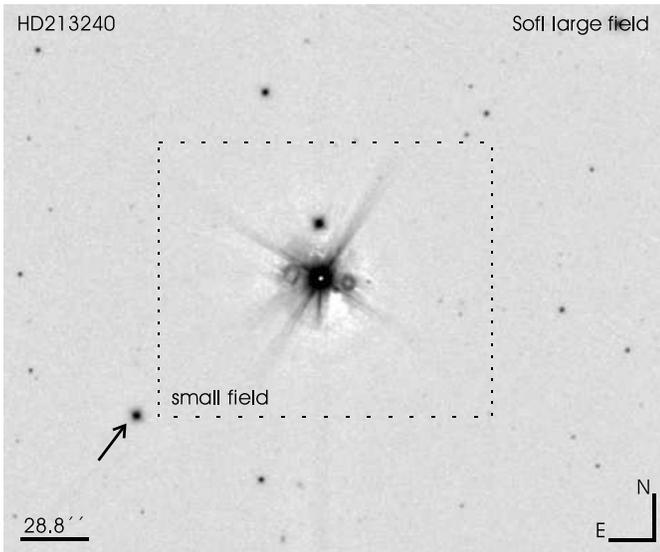}}
\caption{The SofI large field H band image of HD\,213240 taken in July 2004. The total integration
time is 7\,min. Several companion-candidates are detected. An arrow points to the co-moving
companion HD\,213240\,C which was detected by comparing this SofI image with the image from the
2MASS survey.} \label{hd213imagelarge}
\end{figure}

We obtained our first epoch SofI image in December 2002 and a second epoch follow up in July 2004.
Both images are taken as usual in the SofI small field. Only two companion-candidates, both marked
in Fig.\,\ref{hd213imagesmall}, are detected (S/N$\geq$10) with sufficient astrometric accuracy.
The separation of the brighter candidate north of HD\,213240 significantly varies from
22.427$\pm$0.023 in epoch 2002 to 22.684$\pm$0.024\,arcsec. In addition, this companion-candidate
is also listed in the UCAC2 catalogue with a proper motion of $\mu_{Ra}$=65.2\,mas/yr,
$\mu_{Dec}=-10.8$\,mas/yr, hence this object is clearly not co-moving with the exoplanet host star.
For the fainter companion-candidate the separation varies from 20.291$\pm$0.030 in epoch 2002 to
20.022$\pm$0.030\,arcsec in epoch 2004. This change in separation is expected from the Hipparcos
astrometric data of HD\,213240 ($\mu_{Ra}$=-135.16$\pm$0.66\,mas/yr,
$\mu_{Dec}=-194.06\pm$0.47\,mas/yr and $\pi$=24.54$\pm$0.81\,mas), i.e. this faint object is also
clearly not co-moving.

However, HD\,213240 is listed in the Washington Double Star Catalog (WDS) as a binary with a
V=12\,mag companion at the position angle of 284\,$^{\circ}$ and a separation of 19.8\,arcsec
(epoch 1908). Actually, the fainter companion-candidate is located roughly at that position
(PA\,=\,265.68$\pm$0.050\,$^{\circ}$; separation\,=\,20.022$\pm$0.030\,arcsec in July 2004) but as
we have shown above it is clearly not co-moving with HD\,213240. Furthermore, its H magnitude
derived from our H Band images is H=17.517$\pm$0.020\,mag, which is much too faint for a V=12\,mag
object at the distance of HD\,213240. When we extrapolate the proper motion of HD\,213240 back to
the epoch of the WDS entry (1901) and also take into account the proper motion of the brighter
companion-candidate today located north of the exoplanet host star, we derive a position close to
the entry in the WDS for PA and separation (PA\,$\sim$\,281.1$^{\circ}$;
separation\,$\sim$\,20.4\,arcsec). So, it is most likely that this object was denoted as the
secondary in the WDS.

\begin{figure} [ht] \resizebox{\hsize}{!}{\includegraphics{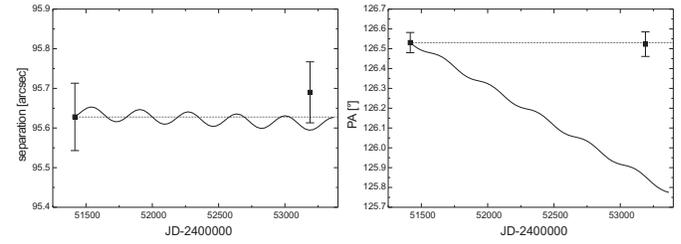}}
\caption{The separation and the position angle between HD\,213240\,C and its primary for the 2MASS
image (epoch 08/99) and our first epoch SofI large field image (epoch 07/04). Separation and
position angle are constant in both observing epochs typical for a co-moving object. Furthermore if
HD\,213240\,C would be non-moving the position angle must significantly change following the
well-known proper and parallactic motion of HD\,213240\,A (see solid lines).} \label{hd213astro}
\end{figure}

In addition to our small field imaging we also observed HD\,213240 with SofI in the large field
mode in July 2004. The given FOV is twice as large as the small field, i.e. 295x295\,arcsec (see
Fig.\,\ref{hd213imagelarge}). Many faint stars can be seen in the image. By comparing our image
with the 2MASS survey from epoch 08/99 we found a co-moving companion. Its separation and position
angle with respect to the primary are constant within the astrometric uncertainty (see
Fig.\,\ref{hd213astro} and Tab.\,\ref{astrometry}). For a non-moving object both parameters must
change due to the proper and parallactic motion of the exoplanet host star. This expected variation
can significantly (8$\sigma$ in position angle) be ruled out for the detected companion (see doted
lines in Fig.\,\ref{hd213astro}). Hence, this companion is clearly co-moving and will be denoted as
HD\,213240\,C in the following, because B is already used by WDS.

\subsection{HD\,16141}

\begin{figure} [ht]
\resizebox{\hsize}{!}{\includegraphics{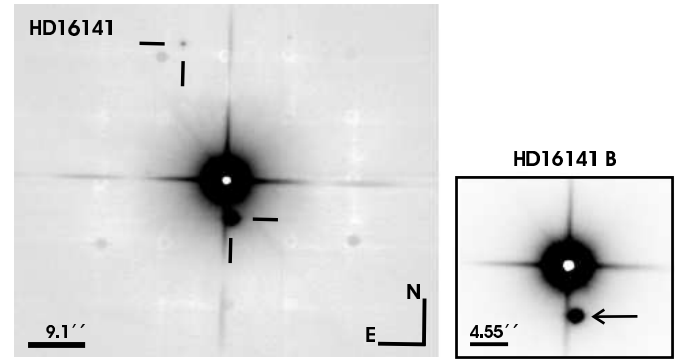}} \caption{Left: The UFTI H Band image obtain
in October 2003. The integration time is 9.6\,min. Two companion-candidates are detected with good
astrometry precision both indicated with markers. Right: The central part of the image zoomed and
rescaled. The newly found co-moving companion HD\,16141\,B is visible $\sim$\,6\,arcsec south of
the exoplanet planet host.} \label{hd161image}
\end{figure}

\begin{figure} [ht] \resizebox{\hsize}{!}{\includegraphics{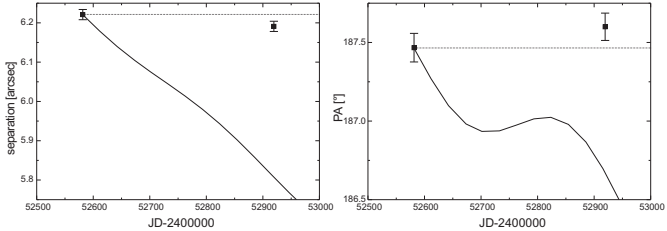}}
\caption{The separation and the position angle of HD\,16141\,B both UFTI/UKIRT images (epoch 11/02
and 10/03). Separation and position angle are constant in both observing epochs as expected for a
co-moving companion and are significantly different to the expected values if HD\,16141\,B was a
non-moving background object using the well-known proper and parallactic motion of the primary (see
solid lines).} \label{hd161astro}
\end{figure}

We observed HD\,16141 for the first time in November 2002, and one year later in October 2003 using
UFTI/UKIRT. Only two companion-candidates are detected with good astrometric precision and are
marked in the left image of Fig.\,\ref{hd161image}. We measured separation and position angle of
the two candidates relative to the exoplanet host star. The separation of the fainter object north
of HD\,16141 increased from 22.394$\pm$0.045\,mas in November 2002 to 22.860$\pm$0.046\,mas in
October 2003 consistent with the proper and parallactic motion of HD\,16141
($\mu_{Ra}$=-156.89$\pm$1.39\,mas/yr, $\mu_{Dec}$=-437.07$\pm$.74\,mas/yr and $\pi$=27.85
$\pm$1.39\,mas). Hence, this is an ordinary background object. The separation and position angle of
the candidate 6\,arcsec south of HD\,16141 is shown in Fig.\,\ref{hd161astro} and in
Table\,\ref{astrometry}. Both separation and position angle are constant in both epochs and are
significantly different (21\,$\sigma$ in separation and 7\,$\sigma$ in position angle) from the
expected data if the candidate was non-moving. Hence, we can conclude that the candidate is clearly
co-moving to HD\,16141, i.e it will be named HD\,16141\,B.

\begin{table} [htb]
\caption{The separations and position angles of all detected companions candidates.}
\begin{center}
\begin{tabular}{c|c|c|c}
\hline\hline
companion & epoch & sep[$arcsec$] & PA [$^{\circ}$]\\
\hline
HD\,114729\,B  & 2MASS 03/99                     &  \,\,\,8.042$\pm$0.085  &  333.151$\pm$0.590\\
small field    & \,\,\,\,\,\,\,NTT 12/02         &  \,\,\,8.043$\pm$0.008  &  333.202$\pm$0.052\\
small field    & \,\,\,\,\,\,\,NTT 06/03         &  \,\,\,8.054$\pm$0.008  &  333.371$\pm$0.052\\
\hline
HD\,196050\,B  & 2MASS 06/00                     &  10.860$\pm$0.085       &  174.360$\pm$0.460\\
small field    & \,\,\,\,\,\,\,NTT 06/03         &  10.880$\pm$0.011       &  174.920$\pm$0.040\\
small field    & \,\,\,\,\,\,\,NTT 06/04         &  10.875$\pm$0.011       &  174.872$\pm$0.040\\
\hline
HD\,213240\,C  & 2MASS 08/99                     &  95.628$\pm$0.085       &  126.531$\pm$0.051\\
large field    & \,\,\,\,\,\,\,NTT 07/04         &  95.690$\pm$0.077       &  126.523$\pm$0.062\\
\hline
HD\,16141\,B   & \,\,\,\,\,\,UFTI 11/02          &  \,\,\,6.221$\pm$0.013  &  187.467$\pm$0.091\\
               & \,\,\,\,\,\,UFTI 10/03          &  \,\,\,6.191$\pm$0.013  &  187.600$\pm$0.087\\
\hline\hline
\end{tabular}
\label{astrometry}
\end{center}
\end{table}

\section{Photometry}

In the previous section we have presented astrometric evidence for companionship of four new
companions. In special cases, apparently co-moving objects may not be physically bound companions,
e.g. in case of a slowly moving foreground or fast moving background object. Hence, it is necessary
to confirm co-moving objects as true companions by e.g. photometry. This section presents the
obtained photometry of all newly found companions derived from our own observations as well as from
the 2MASS point source catalog.

\begin{table} [htb] \caption{2MASS, SofI and UFTI photometry of the newly found companions. The
2MASS photometry of HD\,213240\,C and HD\,196050\,B are consistent with our SofI photometry.
HD\,114729\,B is still slightly contaminated. The closest detected companion HD\,16141\,B is badly
contaminated by more than 1.5\,mag.}
\begin{center}
\begin{tabular}{l|c|c|c|c}
\hline\hline
companion & band & $m_{2MASS}$ & camera & $m_{SofI/UFTI}$\\
\hline
HD\,16141\,B   & J       & \,\,\,9.271$\pm$0.136 & -    &      -               \\
sep$\sim$6\,'' & H       & \,\,\,8.472$\pm$0.325 & UFTI &      10.062$\pm$0.049\\
               & K$_{S}$ & \,\,\,8.766$\pm$0.124 & -    &      -               \\
\hline
HD\,114729\,B  & J       &      10.111$\pm$0.069 & SofI &      10.768$\pm$0.039\\
sep$\sim$8\,'' & H       & \,\,\,9.718$\pm$0.067 & SofI &      10.255$\pm$0.047\\
               & K$_{S}$ & \,\,\,9.517$\pm$0.046 & SofI &      10.008$\pm$0.077\\
\hline
HD\,196050\,B   & J       &      10.619$\pm$0.043 & SofI &      10.777$\pm$0.042\\
sep$\sim$11\,'' & H       &      10.158$\pm$0.037 & SofI &      10.066$\pm$0.057\\
                & K$_{S}$ & \,\,\,9.835$\pm$0.030 & SofI & \,\,\,9.843$\pm$0.079\\
\hline
HD\,213240\,C   & J       &      12.362$\pm$0.025 & -    &      -               \\
sep$\sim$96\,'' & H       &      11.742$\pm$0.026 & SofI &      11.789$\pm$0.090\\
                & K$_{S}$ &      11.465$\pm$0.025& -    &      -               \\
\hline\hline
\end{tabular}
\label{magnitudes}
\end{center}
\end{table}

The angular separation of the detected companions ranges between 96\,arcsec for HD\,213240\,C and
only 6\,arcsec for HD\,16141\,B. The smaller the separation, the stronger the photometric
contamination by light from the nearby bright star. Therefore we obtained the photometry of all
four companions in our H Band images and compared it with the 2MASS photometry. In
Table\,\ref{magnitudes} we have summarized all photometric data. For HD\,213240\,C and
HD\,196050\,B the 2MASS photometry is consistent with the SofI photometry, as expected for such
wide separated companions (more than 10\,arcsec of separation). HD\,114729\,B, with a separation of
8\,arcsec, is slightly contaminated in 2MASS. We obtained additional SofI photometry to confirm
this result also in J and Ks. The 2MASS photometry is about 0.5\,mag brighter in J,H,Ks than our
SofI photometry, as expected for close companions observed with the large 2MASS pixelscale of
1\,arcsec per pixel. Finally, HD\,16141\,B, with a separation of only 6\,arcsec, is badly
contaminated by more than 1.5\,mag in H.  Hence, we do not use 2MASS data for this companion.

\begin{figure} [ht] \resizebox{\hsize}{!}{\includegraphics{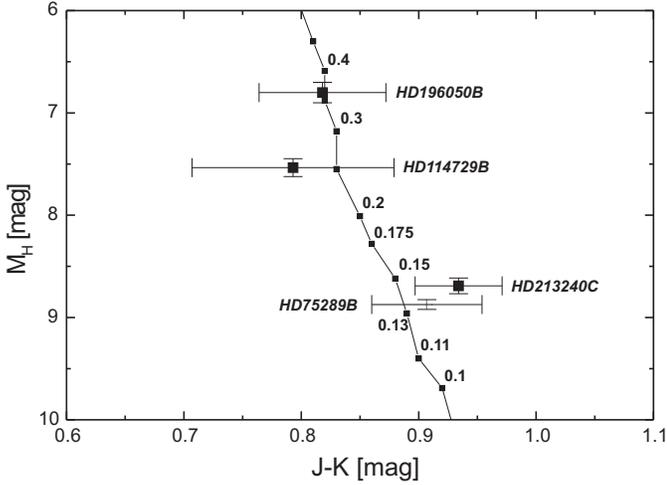}}
\caption{A color-magnitude diagram of all detected companions for which J,H and Ks infrared
magnitudes are known. A isochrone from Baraffe et al. (1998) models is illustrated as solid line
with dots which show the companion-mass in units of solar mass. All companions are consistent with
low-mass stellar objects at the distance of the primaries. In addition we also show HD\,75289\,B
from Mugrauer et al. (2004a). }\label{photo}
\end{figure}

With the transformation of Carpenter (2001) we derive J-K for all companions for which we have
accurate J and Ks magnitudes. We also derived the absolute H magnitude, using the measured apparent
H magnitude and the Hipparcos parallax. We plot all companions in a color-magnitude diagram which
is shown in Fig.\,\ref{photo} (data are summarized in Table\,\ref{photodata}). In addition, the
figure shows the theoretical isochrone for 5\,Gyrs from Baraffe et al. (1998) models. The color and
the derived absolute magnitudes, based on the distance of the primary, are consistent with low-mass
stellar companions. Therefore, we conclude that the photometry is consistent with the companionship
assumption.

\begin{table} [htb]
\caption{J-K color and absolute H magnitudes of the detected companions, shown in
Fig.\,\ref{photo}.}
\begin{center}
\begin{tabular}{c|c|c}
\hline\hline
     companion & J-K [mag] & $M_{H}$ [mag]\\
\hline
HD\,114729\,B   & 0.793$\pm$0.086 & 7.535$\pm$0.088\\
HD\,196050\,B   & 0.818$\pm$0.054 & 6.801$\pm$0.100\\
HD\,213240\,C   & 0.934$\pm$0.037 & 8.691$\pm$0.077\\
HD\,75289\,B    & 0.907$\pm$0.047 & 8.873$\pm$0.048\\
\hline\hline
\end{tabular}
\label{photodata}
\end{center}
\end{table}

\section{Spectroscopy}

Photometry may not be sufficient for confirmation of a co-moving companion because reddened
background giants can be as red as intrinsically red low-mass companions. To confirm that the newly
found companions are really dwarfs we obtained IR spectra of HD\,114729\,B, HD\,196050\,B and
HD\,213240\,C in June 2003 and July 2004 with SofI in spectroscopic mode. We used long slit
spectroscopy with a slit width of one arcsec, and the red grism covering the wavelength range from
1.53 to 2.52\,$\mu$m. The dispersion is 10.22\,\AA~per pixel with an IR HgCdTe detector in the
large field mode (288\,mas pixel scale) providing a resolving power of $\lambda/\Delta \lambda
\approx$\,588.

Background subtraction was obtained by nodding between two positions along the slit, as well as by
a small jitter around those two positions, to avoid individual pixels always seeing the same part
of the sky.  All images were flat fielded with a standard dome flat and wavelength calibrated with
a Xe lamp. After flatfielding, all individual spectra, each with a total integration time of
1\,min, were averaged. We used standard IRAF routines for background subtraction, flat fielding and
averaging all individual spectra.

\begin{figure} [ht] \resizebox{\hsize}{!}{\includegraphics{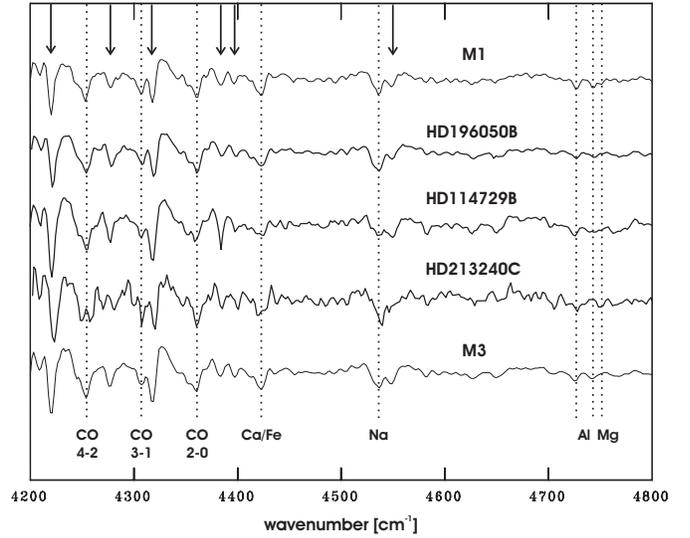}} \caption{K band SofI spectra of
the detected companions HD\,114729\,B, HD\,196050\,B and HD\,213240\,C. Furthermore we show
comparison spectra of spectral type M1V and M3V. The arrow indicate telluric features.}
\label{spectra}
\end{figure}

Fig.\,\ref{spectra} shows the normalized K band spectra of the detected companions and two
comparison spectra (HIP\,80268 B-V=1.458 M1V, HIP5496 B-V=1.568 M3V). The strongest lines in the
spectra are from molecular bands of the first CO overtone, extending from 4360\,cm$^{-1}$ to the
low frequency side of the spectrum. Strong atomic features of Ca/Fe (4415\,cm$^{-1}$) and Na
(4530\,cm$^{-1}$) are visible, as well as weak lines of Al at 4724 and 4740\,cm$^{-1}$ and Mg at
4747\,cm$^{-1}$. Hence all three companions are of spectral type early M, which is also consistent
with the measured colors. Due to CO bands which are only slightly stronger than the atomic features
of Na and Ca/Fe we can conclude that all three companions are dwarfs, hence luminosity class V.

\section{Discussion}

\subsection{The primaries and their planets}

\subsubsection{HD\,16141}

HD\,16141 is a slow rotating chromospherically inactive G5IV star, located in the constellation
Cetus. With the Keck High-Resolution Echelle Spectrograph Marcy et al. (2000) found an exoplanet
($msin(i)$\,=\,0.215\,$M_{Jup}$) orbiting its parent star on a 0.35\,AU eccentric orbit (e=0.28)
with an orbital period of only 0.21\,yr. Laws et al. (2003) determined that the exoplanet host star
is between 3.2 and 6.7\,Gyr old and has a mass of 1.22\,$M_{\sun}$.

\subsubsection{HD\,114729}

Butler et al. (2003) reported the detection of an exoplanet orbiting the star HD\,114729, a result
of the Keck Precision Doppler Survey. The planet has a minimum-mass $msini(i)$\,=\,0.84\,$M_{Jup}$
and revolves around its host star on a 3.11\,yr eccentric orbit, with a semi-major-axis ($a$) of
2.08\,AU and an eccentricity ($e$) of 0.32. The parent star is located in Centaurus, listed as G3V
by SIMBAD and G0V by Hipparcos. The latter classification was assigned by Butler et al.(2003)
derived from their Keck spectra. The star is chromospherically quiet and Butler et al.(2003)
estimate its age to be approximately 6\,Gyrs.

\subsubsection{HD\,196050}

HD\,196050 is a G3V star located in the southern constellation Pavo. The discovery of an exoplanet
was reported firstly by Jones et al. (2002) in the Anglo-Australian Planet Search (AAPS) and
confirmed later by Mayor et al. (2004) in the CORALIE planet-search programm. The detected
exoplanet ($msin(i)$\,=\,2.8\,$M_{Jup}$) revolves its host star on an eccentric 3.56\,yr orbit
(e\,=\,0.19; a\,=\,2.4\,AU). The parent star is chromospherically quiescent and is not detected to
be variable by Hipparcos. It is used as an infrared spectroscopic standard by the ESO/NTT team on
La Silla. The mass and age of HD\,196050 was determined by Mayor et al. (2004) to be
1.1\,$M_{\sun}$ and 1.6\,Gyrs, respectively.

\subsubsection{HD\,213240}

HD\,213240 is a G0V star located in the southern constellation Grus. Its mass is 1.22\,$M_{\sun}$
and its age lies between 2.7 and 4.6\,Gyrs, with the stellar parameters derived by Santos et al.
(2001). The same group reported the detection of an exoplanet with a minimum mass of 4.5\,$M_{Jup}$
which revolves around HD\,213240 on a 2.61\,yr eccentric (a=2.03\,AU, e=0.45) orbit. Although the
minimum mass of the close companion is high compared to most of the detected exoplanets, Santos et
al. (2001) noted that the probability that the companion is actually a brown dwarf and not a planet
is quite low, only around 5\%, and thus the planetary assumption seems the most probable.

\subsection{Statistical properties of exoplanets in binaries}

\begin{figure} [hbt]
\includegraphics[height=15cm]{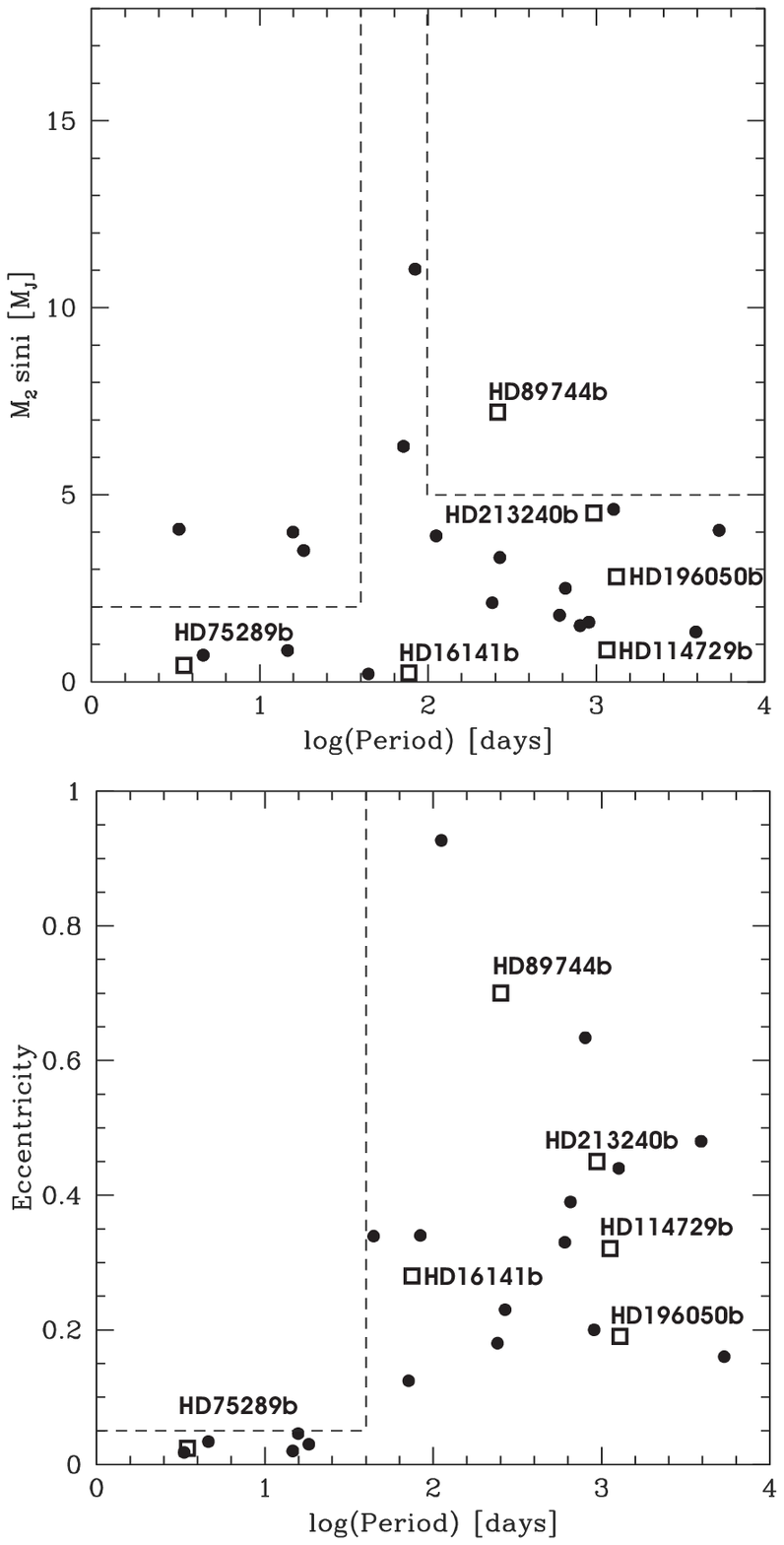} \caption{The updated mass-period (upper panel)
and eccentricity-period distribution (lower panel) of Eggenberger et al. (2004). All new
binary-star planets are marked with black squares and all binary-star planets from Eggenberger et
al. (2004) are shown with dark dots. HD\,213240\,b, HD\,196050\,b, and HD\,114729\,b exhibit
minimum masses smaller 5\,$M_{Jup}$ comparable to all other binary-star planets found with periods
longer than 100 days. However with $Msin(i)=7.2\,M_{Jup}$ HD\,89744\,b is the first known exception
from that rule. HD\,75289\,b exhibits a low eccentricity consistent with all other known
binary-star planets on orbits shorter 40 days. The eccentricities of HD\,213240\,b, HD\,196050\,b,
HD\,114729\,b and HD\,89744\,b are widely spread from 0.19 up to 0.7, consistent with the
eccentricity distribution of other long periodic binary-star as well as single-star planets. The
eccentricity distribution of both exoplanet populations seems to be similar for orbital periods
longer than 100 days.}\label{planets}
\end{figure}

We can group all exoplanets in two populations, namely exoplanets found in binaries (binary-star
planets) and planets orbiting a single star (single-star planets). So far we have detected 6
further binaries among the exoplanet host stars in our survey, i.e. the sample size of binaries
among exoplanet host stars is extended by 40\,\% compared to Eggenberger et al. (2004) who describe
differences in the properties (period, mass, eccentricity) of the two exoplanet populations.

In the mass-period diagram Zucker \& Mazeh (2002) pointed out that for periods shorter than 40 days
the most massive planets are binary-star planets and all known single-star planets have
minimum-masses $msin(i)<2M_{Jup}$. Eggenberger et al. (2004) count 3 binary-star planets with
$msin(i)>2M_{Jup}$ and only 2 binary-star planets but 20 single-star planets with
$msin(i)<2M_{Jup}$. Selection effects are taken into account by discarding from the count
exoplanets with $K<15m/s$. If no difference in the mass-period distribution of the both exoplanet
populations is assumed, the probability for the given binary planet distribution is only 0.44\,\%

In the eccentricity-period diagram both populations seems to be different as well. Eggenberger et
al. (2004) count 5 binary-star planets with $e<0.05$ and single-star planets appear to be more or
less equally distributed, 9 with $e<0.05$ and 11 with $e>0.05$. They conclude that with a
probability of 3.77\,\% the given binary-star planet distribution can be explained assuming that
both populations are identical.

Among the new binary-star planets, HD\,75289\,b ($msini(i)=0.42M_{Jup}$, a=0.046\,AU, e=0.024)
revolves its host stars with an orbital period of less than 40 days, hence modifies the mass-period
and eccentricity-period diagram of the close binary-star planets (see Fig.\,\ref{planets}).The
derived probabilities are updated to 0.87\,\% \footnote{Hypergeometric distribution:
$\left(\begin{array}{c}22\\3\\\end{array}\right)\left(\begin{array}{c}3\\3\\\end{array}\right)\,/\,\left(\begin{array}{c}25\\6\\\end{array}\right)$=0.87\,\%}
in the mass-period diagram and 1.70\,\% \footnote{Hypergeometric distribution:
$\left(\begin{array}{c}14\\6\\\end{array}\right)\left(\begin{array}{c}11\\0\\\end{array}\right)\,/\,\left(\begin{array}{c}25\\6\\\end{array}\right)$=1.70\,\%}
in the eccentricity-period diagram, respectively.

The fact that the mass of the planet around HD\,75289 is less than 2\,$M_{Jup}$ slightly decreases
the statistical significance of the {\it mass} difference between the short-period single-star and
binary-star planets. However, the significance remains still high. On the other hand the difference
in the eccentricity distribution becomes slightly larger, supporting the conjecture that the
multiplicity of the exoplanet host stars indeed influences either the planet formation process or
the long-term orbital evolution of close exoplanets. The observed differences in the mass- and
eccentricity-period diagrams are expected if a migration process took place in the history of the
planetary systems. The multiplicity of the host star should speed up the migration process,
yielding more massive short period planets on low-eccentric orbits compared to exoplanets in single
star systems (see e.g. Kley 2000).

Eggenberger et al. (2004) also describe a difference in the mass-period distribution between
single-star and binary-star planets with orbital periods longer than 100 days. They count 10
binary-star planets with $msin(i)<5M_{Jup}$ but no more massive binary-star planets. The retracted
planet HD\,219542\,b (Desidera et al. 2003) was already excluded there because of K=13\,m/s and is
not shown in Fig.\,\ref{planets}. In contrast there are 46 single-star planets with
$msin(i)<5M_{Jup}$ and 21 with $msin(i)>5M_{Jup}$. Assuming again the same distribution for both
populations the probability of finding 10 planets with $msin(i)<5M_{Jup}$ but 0 with with
$msin(i)>5M_{Jup}$ is 3.25\,\%.

The new binary-star planets HD\,213240\,b, HD\,196050\,b and HD\,114729\,b are all located in this
region of the mass-period diagram, hence now there are 13 long period binary planets known with
$msin(i)<5\,M_{Jup}$. HD\,89744\,b ($msini(i)$=7.2\,$M_{Jup}$, a=0.88\,AU, e=0.70) is the first
known long-period massive planet, found by us to reside in a binary system. With updated statistics
the derived probability of having the given masses of the binary-star planets coming from the same
distribution for both exoplanet populations increases from 3.25\,\% to 4.73\,\%
\footnote{Hypergeometric distribution:
$\left(\begin{array}{c}56\\13\\\end{array}\right)\left(\begin{array}{c}21\\1\\\end{array}\right)\,/\,\left(\begin{array}{c}77\\14\\\end{array}\right)$=4.73\,\%},
i.e. the significance of the reported statistical difference is weakened. For orbital periods
longer than 100 days, binary- and single-star planets display the same distributions, as was
already apparent from the eccentricity-period distribution. Therefore, it seems as if the host star
multiplicity does not affect the orbital properties of the long-period exoplanets.

\subsection{The newly detected secondaries}

All newly found companions presented in this work are clearly co-moving with their primaries, the
exoplanet host stars. For HD\,114729\,B, HD\,196050\,B and HD\,213240\,C we have also obtained
J,H,K$_{s}$ photometry as well as K band spectra. The derived spectral types and the measured
infrared colors and magnitudes are consistent with low-mass stellar objects at the well-known
distances of the exoplanet host stars. Hence, companionship is also confirmed by photometry and
spectroscopy. For HD\,16141 we only have astrometric evidence for companionship. However the chance
of finding a co-moving object at a separation of only 6\,arcsec at a galactic latitude of
$\sim$-56$^{\circ}$ is negligible. To confirm our detection, follow-up spectroscopy is planned for
this companion.

We derive the companion masses from the given infrared colors and Hipparcos parallaxes of the
primaries using the Baraffe et al. (1998) isochrones for a system age of 5\,Gyrs. The age
uncertainty of the primaries is not very important, because magnitudes of low-mass stellar
companions are very stable for an age range between 1 and 10\,Gyrs. In Table\,\ref{info} we
summarize the derived parameters of the detected co-moving companions.

\begin{table} [htb]
\caption{Projected separation and derived companion masses.}
\begin{center}
\begin{tabular}{c|c|c}
\hline\hline
     companion & proj. separation [AU]& mass [$M_{\sun}$]\\
\hline
HD\,16141\,B   & \,\,\,223$\pm$11 & 0.286$\pm$0.017\\
HD\,114729\,B  & \,\,\,282$\pm$10 & 0.253$\pm$0.011\\
HD\,196050\,B  & \,\,\,511$\pm$22 & 0.363$\pm$0.018\\
HD\,213240\,C  & 3898$\pm$129     & 0.146$\pm$0.005\\
\hline\hline
\end{tabular}
\label{info}
\end{center}
\end{table}

With the derived companion masses, the primary mass ($\sim$1$M_{\sun}$) and the companion
separations we can compute the expected RV variation of the primary induced by the presence of the
wide companions. The maximal yearly variation of the RV is $\sim$1\,m/s for HD\,16141, the closest
detected companion, being probably detectable as linear trend in the RV after a decade of precise
radial velocity measurements.

\begin{figure} [ht]
\resizebox{\hsize}{!}{\includegraphics{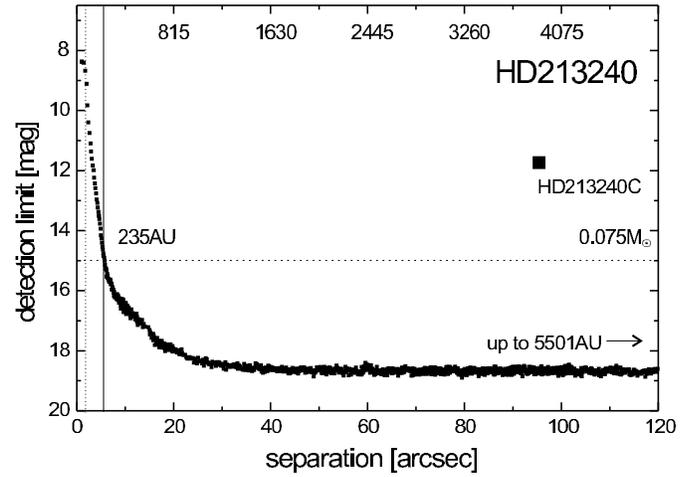}}\caption{Detection limit of SofI large
field in H band for a range of separation given in arcsec at the bottom and as projected separation
in AU at the top. At $\sim$\,1\,arcsec saturation occurs (doted line). All stellar companions
(m\,$\ge$\,0.075\,$M_{\sun}$) are detectable beyond the distance illustrated by the straight
line.}\label{dynamic1}
\end{figure}

\begin{figure} [ht]
\resizebox{\hsize}{!}{\includegraphics{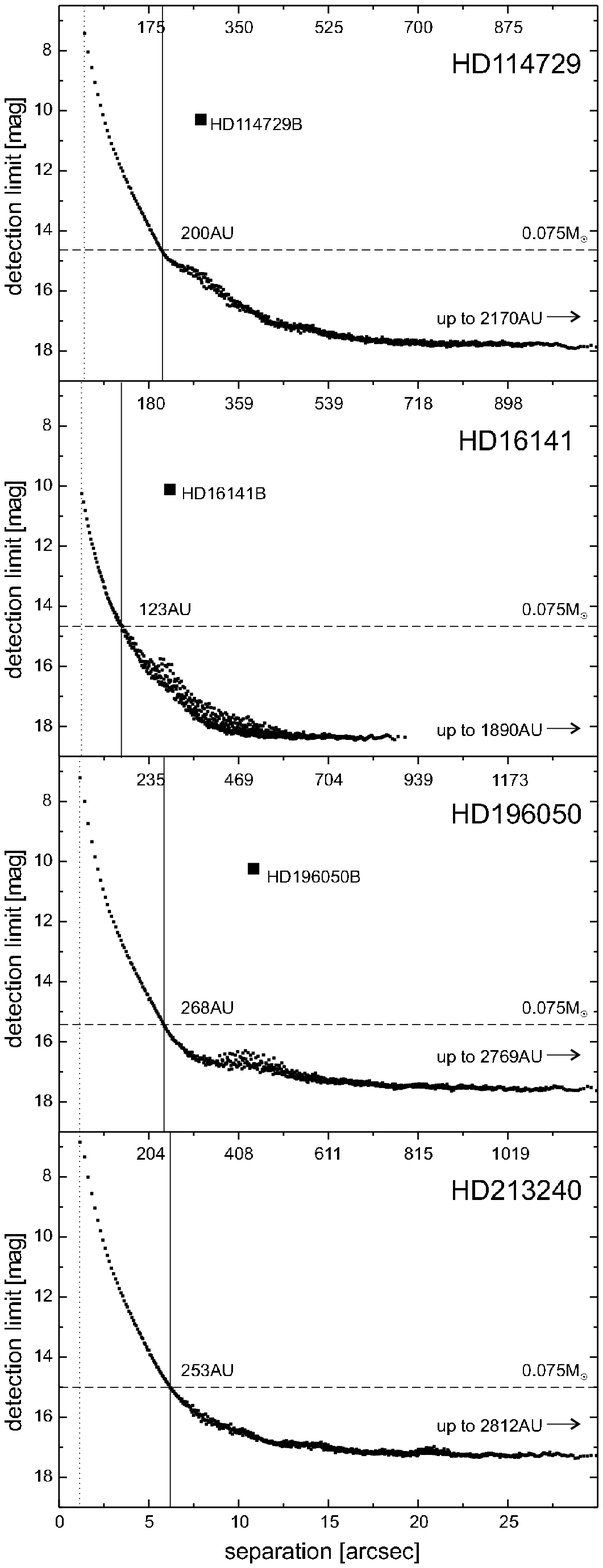}}\caption{Detection limits of SofI small field
and UFTI images in H band for a range of separations given in arcsec at the bottom and as projected
separation in AU at the top of each individual plot. At $\sim$\,1\,arcsec saturation occurs (doted
line). All stellar companions (m\,$\ge$\,0.075\,$M_{\sun}$) are detectable beyond the distance
illustrated by the straight lines in each individual plot.}\label{dynamic2}
\end{figure}

Figure\ref{dynamic1} and Fig.\ref{dynamic2} shows the achieved SofI and UFTI detection limits
(S/N=10), which reach $\sim$18\,mag in H, and therefore enables the detection of substellar
companions down to M$_{H}$\,$\sim$\,15\,mag, which is translated to a mass of $\sim$\,60\,M$_{Jup}$
according to the Baraffe et al. (2003) models. Objects at distances of up to $\sim$\,60\,arcsec
were observed twice but no further co-moving companion could be identified. Further stellar
companions (m$\ge$75\,M$_{Jup}$) can be ruled out for a projected separation from $\sim$200\,AU up
to $\sim$2400\,AU.

\acknowledgements {We would like to thank the technical staff of the ESO NTT and Joint Astronomy
Center for help and assistance in carrying out the observations and A.~Szameit who carried out some
of the observations. We made use of the 2MASS public data releases as well as the Simbad database
operated at the Observatoire Strasbourg. T.M. thanks the Israel Science Foundation for a support
through grant no.03/233.}

{}
\end{document}